\documentclass[aps,pre,superscriptaddress,notitlepage,10pt]{revtex4-1}

\usepackage{amsmath,amssymb,amsfonts}
\usepackage{graphicx}
\usepackage[dvipsnames]{xcolor}
\usepackage{comment}
\usepackage[overload]{empheq}
\usepackage{upgreek}
\usepackage{lmodern,pdftexcmds}
\usepackage[sort&compress]{natbib}
\usepackage{hyperref}
\usepackage{soul}
\usepackage[normalem]{ulem}
\usepackage{enumerate}

\hypersetup{
    bookmarksopen=true,
    unicode=false,          
    pdftoolbar=true,        
    pdfmenubar=true,        
    pdffitwindow=false,     
    pdfstartview={Fit},    
    pdftitle={},    
    pdfauthor={},     
    colorlinks=true,       
    linkcolor=black,          
    citecolor=black,        
    urlcolor=black,           
    pdfborder={0 0 0}
}

\newcommand{\beq}{\begin{equation}}
\newcommand{\eeq}{\end{equation}}
\newcommand{\beqa}{\begin{eqnarray}}
\newcommand{\eeqa}{\end{eqnarray}}

\providecommand*{\pp}[3][]{\frac{\partial^{#1}#2}{\partial #3^{#1}}}
\providecommand*{\dd}[3][]{\frac{\mathrm{d}^{#1}#2}{\mathrm{d} #3^{#1}}}

\newcommand{\qmbox}[1]{\quad \mbox{#1} \quad}

\providecommand*{\rmd}{\mathrm{d}}

\begin{document}

\preprint{}

\title{Representative subsampling of sedimenting blood}



\author{Bhargav Rallabandi}
\email{bhargav@engr.ucr.edu}
\affiliation{Department of Mechanical Engineering, University of California, Riverside, California 92521, USA }
\affiliation{Department of Mechanical and Aerospace Engineering, Princeton University, Princeton, New Jersey 08544, USA}

\author{Janine K. Nunes}
\affiliation{Department of Mechanical and Aerospace Engineering, Princeton University, Princeton, New Jersey 08544, USA}

\author{Antonio Perazzo}
\affiliation{Department of Mechanical and Aerospace Engineering, Princeton University, Princeton, New Jersey 08544, USA}

\author{Sergey Gershtein}
\affiliation{Abbott Point of Care, Princeton, New Jersey 08540, USA}

\author{Howard A. Stone}
\email{hastone@princeton.edu \\}
\affiliation{Department of Mechanical and Aerospace Engineering, Princeton University, Princeton, New Jersey 08544, USA}

\date{\today}

\begin{abstract}
It is often necessary to extract a small amount of a suspension, such as blood, from a larger sample of the same material for the purposes of diagnostics, testing or imaging. A practical challenge is that blood sediments noticeably on the time scale of a few minutes, making a representative subsampling of the original sample challenging. Guided by experimental data, we develop a Kynch sedimentation model to discuss design considerations that ensure a representative subsampling of blood for the entire range of physiologically relevant hematocrit over a specified time of interest. Additionally, we show that this design may be modified to exploit the sedimentation and subsample either higher or lower hematocrit relative to that of the original sample. Thus, our method provides a simple tool to either concentrate or dilute small quantities of blood or other sedimenting suspensions.
\end{abstract}


\maketitle 
\section{Introduction}
Diagnostic assays are an integral part of clinical practice. Whole blood is one of the most common sample types used in diagnostics \citep{cui15_sampleprep_review} and is an important material for point-of-care applications as it can be collected relatively easily in small volumes. Blood is made up primarily of blood plasma (about 55 \% by volume). The remainder is cellular content, the vast majority of which comprises erythrocytes (red blood cells), which outnumber both the slightly larger white blood cells and the smaller platelets that together make up only about 1\% of blood volume. Blood plasma comprises about 92\% by weight of water with the remaining consisting of a dissolved proteins (around 7\%), sugars and salts.

The demand for sample analysis at low cost and high accuracy has made microfluidic techniques increasingly popular in recent years. Much attention has been focused on separating plasma from whole blood for diagnostics \citep{ker13_microcalesep_review}, in particular because traditional centrifugal fractionation of blood can be labor intensive and expensive. Recent studies have sought to address these challenges by developing microfluidic plasma filtration devices using a host of techniques including membrane-based \citep{hom12_plasmasep,lu18_separation_paper} or channel-based microfilters \citep{fai06_geometrical_separation,wu12_blood_filtration_microfilter, kuo18_channelfilter}. Other have used external fields to apply forces to cells and separate them from plasma, including electrokinetic \citep{min02_electrokinetic}, magnetic \citep{jun08_magnetic, kim12_removal, tas15_levitation_magnet}, acoustic \citep{len09_acoustic}, and inertial \citep{mac10_inertial_filtration} forces. Notable within the category of separation devices using external fields are those that exploit the gravitational sedimentation of blood cells \citep{dim11_bloodanalysis}, with some designs additionally making use of fluid flow \citep{zha12_gravitational}.  Others have developed separation devices on centrifugal microfluidic chips by incorporating microvalves \citep{li10_separation} or by employing centrifugal forces in curved or branched microchannels  \citep{zha08_separation}. There has also been an interest in cell-scale processes \citep{fre14_annurev,tom14_biomechanical}, including the effects of erythrocyte morphology and deformability on blood rheology \citep{for11_multiscale_RBC_dynamics,lan16_morphology_viscosity}, and the apparent dependence of blood viscosity on the radius of the capillary (the F\r{a}hr{\ae}us--Lindqvist effect) \citep{sec17_blood_annurev}. Cell deformability has also been used to develop microfluidic cell sorters \citep{fai06_geometrical_separation, guo14_deformability_microfluidic} and is important to the life cycle of erythrocytes \citep{piv16_spleen}.

A typical blood sample in diagnostics (extracted, for example, by a pin-prick) has a volume of about $50$ $\mu$L. In contrast, the volume that can be contained in a microfluidic device is much smaller (typically a few $\mu$L). Thus, for most diagnostic techniques using microfluidics, it is usually only necessary for a small amount of the original sample to be processed, i.e. the original sample must be ``subsampled'' for the diagnostic assay. The cellular constituents of blood are denser than the surrounding plasma and will therefore sediment over time due to an external force such as gravity or a centrifugal force. For a typical cell volume fraction -- or hematocrit -- of about 45 \%, the sedimentation speeds of erythrocytes under gravity are about $0.2$--$0.3$ mm/min. Sedimentation rates are faster for lower cell fractions \citep{mil83_sediment, jac87_sed_hem}. Thus, with processing times on the order of several minutes, sedimentation is appreciable in millimeter-sized geometries: over time, the top of the sample continuously becomes more dilute, while cells aggregate near the bottom of the sample. As discussed above and in previous studies \citep{dim11_bloodanalysis, zha12_gravitational,sun12_blood_sep_twophase_sedimentation}, this is a useful property  for separating plasma from cells. It is, however, counterproductive if the goal is to maintain the composition of the sample throughout the sampling time.

Here, we present experiments and a corresponding model for the sedimentation of erythrocytes in a sample blood volume. Our primary goal is to use the predictions of this model to identify design considerations that extract a representative subsample in spite of the continuous sedimentation of the blood sample. We identify the region of the design space, which involves the device geometry and the sampling (or filling) time, for which representative subsampling is possible. Furthermore, we are able to tune the design parameters so that the collected subsample has a prescribed concentration of cells different from that of the original sample, allowing us to design subsampling protocols to collect either a more dilute or more concentrated extract than the original sample. The remainder of the paper is organized as follows: Sec. \ref{SecModel} introduces the experimental procedure and a corresponding model that quantifies blood sedimentation. Section \ref{SecSubsampling} discusses solutions of the model in the context of extracting either representative or non-representative subsamples. In Sec. \ref{SecSideChannel} we discuss the process of subsampling by a microfluidic channel (or network of channels) throughout the sedimentation process, after which we conclude in Sec. \ref{SecConcl}.

\section{Sedimentation and design considerations}  \label{SecModel}

\begin{figure}[t!]
    \includegraphics[scale=1]{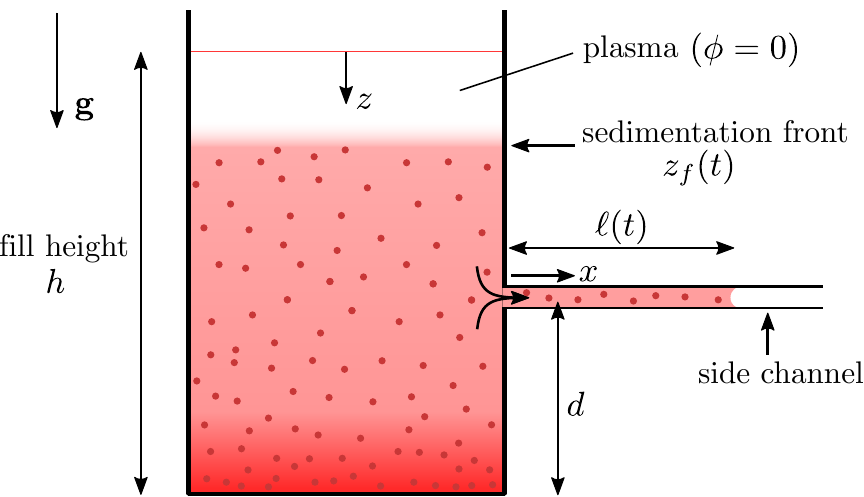}
\caption{A sketch of the setup showing a reservoir filled with a blood sample up to a height $h$. Connected to the reservoir at a height $d$ is a narrow side channel that continuously draws a small volume of fluid from the reservoir; the fluid collected within the side channel forms the subsample. Over time the suspension in the reservoir sediments and develops a moving front $z = z_f(t)$, such that the reservoir contains only plasma (cell volume fraction $\phi = 0$) between the front and the fill line $(z = 0)$. As the sedimentation proceeds, the opening of the side channel is presented with suspension properties that vary in time. An applied horizontal pressure gradient pumps fluid into the side channel, causing it to be filled to a length $\ell(t)$ that increases with time.}
\label{FigSketch}
\end{figure}
We are interested in subsampling a volume $V_s$ of blood from a larger sample of volume $V > V_s$. The original sample has a volume fraction of red blood cells (erythrocytes) $\phi \leq 1$, often referred to as the hematocrit (and expressed as a percentage). The volume fraction is typically around $0.45$, though it may be much higher (up to $0.7$) or lower (as low as $0.1$) depending on physiological or pathological conditions \citep{bri99_clinica_ESR}. White blood cells and platelets together make up only about 1\% of blood volume and so their contribution to the dynamics discussed below will be neglected.

Here we discuss design criteria that ensure a representative subsampling for the entire range of physiologically relevant hematocrit $(0<\phi\lesssim 0.75)$. A sketch of the gravitational sedimentation setup considered in this work is shown in Fig. \ref{FigSketch}. We consider a reservoir that is filled with the original sample volume $V$, connected to which is a side channel that continuously draws in suspension from the reservoir over time. The fill height of the sample in the reservoir is $h = V/A$, where $A$ is the constant area of cross-section of the main container. The side channel is located at a distance $d$ from the bottom of the reservoir. We are interested in the properties of the fluid drawn into the side channel -- the subsample -- relative to those of the original sample placed in the reservoir. 

\subsection{Sedimentation Experiments}
The sedimentation rate $v$ of erythrocytes depends on several factors including the mean hematocrit \citep{rou30_sediment}, the protein content of the plasma \citep{rop39_sed_plasma}, the age of the sample, pathological and physiological conditions, and the presence of additives \citep{bri99_clinica_ESR}. Here, we focus on the dependence of $v$ on $\phi$, which is particularly important as the local hematocrit of a given sample of blood changes during the sedimentation process. Despite the many variables controlling the sedimentation rate it is agreed that the sedimentation rate decreases with hematocrit, provided the geometry is large compared with the size of the cell and cell-aggregates \citep{fab87_aggregation_sedimentation}. 

We quantify the dependence of sedimentation rate on hematocrit by direct measurements. Unspun whole blood with tri-potassium ethylenediaminetetraacetic acid (EDTA, an anti-coagulant) was received from Biological Specialty Corporation (Colmar, PA). The blood was centrifuged at 3000 rpm for 10 minutes to separate the red blood cells from the plasma; the blood, as received, was 45\% hematocrit. Then, 10 mL each of different hematocrit samples were prepared by mixing different volume ratios of cells and plasma in 15 mL centrifuge tubes (inner diameter = 1.4 cm). Prior to sedimentation, the tubes were inverted repeatedly and gently by hand to mix the suspensions, and then placed upright in Styrofoam racks. Images were captured every minute during sedimentation (Nikon D5100, Camera Control Pro). The images were analyzed using ImageJ software \citep{sch12_ImageJ} to track the location of the sedimentation front over time (see Fig. \ref{FigSketch}). Consistent with previous studies \citep{rou30_sediment,fab87_aggregation_sedimentation} we observe that the sedimentation front moves at a roughly constant speed over a range of times. We determine the maximum sedimentation rate for each hematocrit by analyzing this part of the displacement versus time curve (see \citep{rou30_sediment}).  Sedimentation experiments were conducted simultaneously in four tubes per hematocrit at room temperature (21--22$^\circ$C).

\begin{figure}
    \centering
    \includegraphics[scale=1]{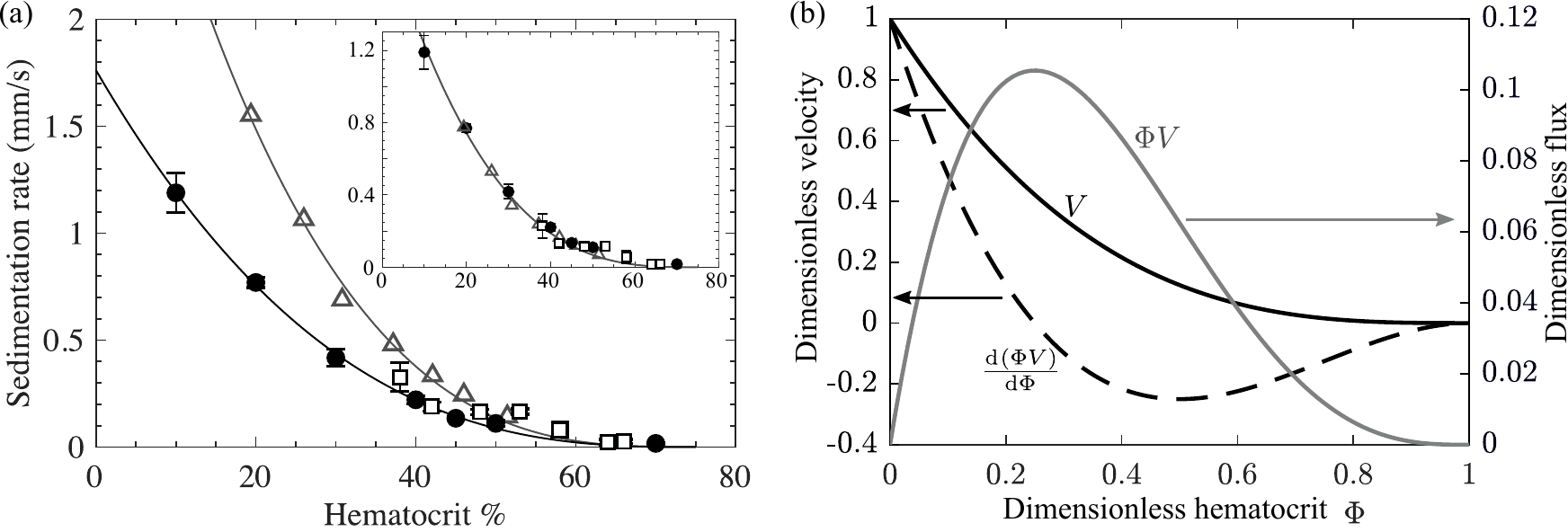}
\caption{(a) Experimentally measured sedimentation rates as a function of the hematocrit: filled circles and open squares indicate two sets of present experimental results (whole blood samples from two different anonymous donors) and triangles are the data of \citep{rou30_sediment}. Error bars indicate one standard deviation. Curves are best fits of the sedimentation law \eqref{SedimentationLaw} with $\phi_m = 0.75$; a fit to the current experiments (circles) yields $v_0 \approx 1.76$ mm/min, $k \approx 2.75$ and the fit to the data of \citep{rou30_sediment} results in $v_0 \approx 3.81$ mm/min, $k \approx 3.03$. The inset plots the same data but with the sedimentation rates represented by squares and triangles being scaled by a uniform factors of $0.7$ and $0.5$, respectively, showing that all three data sets are similar up to scale. (b) Dimensionless sedimentation velocity $V(\Phi) = (1-\Phi)^k$ (solid; left axis), flux $J(\Phi) = \Phi V$ (solid; right axis) and wave speed $U(\Phi) = \dd{(\Phi V)}{\Phi}$ (solid; left axis) with  $k = 3$. The flux is maximum when $\Phi = (1 + k)^{-1}$.}
\label{FigFlux}
\end{figure}

Fig. \ref{FigFlux}(a) plots sedimentation speed as a function of the sample hematocrit, extracted from our measurements (solid circles), alongside the measurements of Rourke and Ernstene \citep{rou30_sediment} (triangles). In both cases, we find that the data are well fit by a sedimentation law of the form 
\begin{equation} \label{SedimentationLaw}
    v(\phi) = v_0 \left(1 - \frac{\phi}{\phi_m}\right)^k,
\end{equation}
where $v_0$ is the sedimentation speed in the dilute limit, $\phi_m$ is the hematocrit at which no sedimentation can be observed and $k$ is a dimensionless exponent. With $\phi_m = 0.75$, a best fit to our data gives $v_0 \approx 1.76$ mm/min and the exponent $k = 2.75$, while a fit to the data of \citep{rou30_sediment} yields $v_0 \approx 3.81$ mm/min  and $k \approx 3.03$. 

To further validate our data, we performed separate measurements with a different whole blood sample (obtained from a different anonymous donor) of volume $\approx 60\,\mu$L in a narrow capillary tube (inner diameter of 1.1--1.2 mm). Sedimentation rates were measured using a similar procedure as the one described above. The results of these measurements are indicated as open squares in Fig. \ref{FigFlux}(a) and are largely consistent with the data obtained in the wider tubes (circles). Both datasets corresponding to the present experiments suggest slower sedimentation compared with \citep{rou30_sediment}, which we speculate may be due to factors such as differing physiological or pathological conditions that we do not control (cf. \citep{fab87_aggregation_sedimentation}). Interestingly, within the range of measurement, the data sets differ by a scale factor between $0.5$ and $0.7$ but are nearly identical in shape, as indicated in the inset of Fig. \ref{FigFlux}(a). As noted above, the detailed values of $v_0$ and $k$ may depend on factors other than the hematocrit and will not be directly relevant to the discussion below. The more important feature is that \eqref{SedimentationLaw} provides a good representation of the dependence of sedimentation rate on hematocrit.  

Equation \eqref{SedimentationLaw} may also interpreted in terms of a Krieger--Dougherty law \citep{kri59_suspensions} for the viscosity of a suspension $\eta(\phi) = \eta_0 \left(1 - \frac{\phi}{\phi_m}\right)^{-k}$ by writing $v(\phi)=  \frac{g(\Delta m)}{6 \pi a \eta(\phi)}$, where $a$ is a length scale characterizing Stokes drag on an erythrocyte, $\Delta m$ is its buoyant mass, $g$ is the gravitational acceleration, and $\eta_0$ is the plasma viscosity. We recognize that this description is only approximate since blood is a complex fluid that is shear-thinning, slightly viscoelastic and has a small yield stress (1--10 mPa) \citep{hor18_loas}. However, its rheology is often well described by a Casson law \citep{apo14_rheology_shear,tom16_casson}, which, in the limit of a small yield stress, reduces to an Newtonian effective-viscosity description such as the one used above.

\subsection{Sedimentation model}
We describe the sedimentation of the suspension using a one-dimensional Kynch model \citep{kyn52_sedimentation}, where the sedimentation speed depends only on the local volume fraction of cells, which in turn evolves in space and time. We take $z = 0$ at the top of the sample and $z = h$ at the bottom of the reservoir (Fig. \ref{FigSketch}). Writing $\phi = \phi(z,t)$, conservation of cell number in the reservoir containing the original sample can be expressed as  
\begin{align}
\pp{\phi}{t} + \pp{(\phi v(\phi))}{z} = 0.
\end{align}
Note that we have neglected the loss of cells (and plasma) into the subsampling side channels, which implicitly assumes that the subsample volume is much smaller than that of the original sample ($V_s \ll V$). We also assume that the side channel is sufficiently narrow to be able to sample the container at a single height $d$ from the bottom, and that the flow into the side channel does not influence the sedimentation in the main container. This loss of sample into the side channel may be modeled within the present framework by a point sink of material placed at a distance $z = h-d$, though we will neglect it in our discussion below. 
For a one-dimensional model, the physically relevant boundary condition is that of zero cell flux $\phi v(\phi)$ at the top and bottom boundaries.

 We also assume that the reservoir is instantaneously filled with a uniform sample, $\phi(0<z<h,\, t=0) = \phi_0$. Introducing dimensionless variables $Z = z/h$, $T = t v_0/h$, $\Phi = \phi/\phi_m$, and defining
\begin{align} \label{Vnondim}
V(\Phi) = \frac{v(\phi)}{v_0} = (1-\Phi)^{k},
\end{align}
sedimentation in the reservoir is governed by
\begin{subequations} \label{SedGE}
\begin{align}
    \pp{\Phi}{T} + \pp{(\Phi V(\Phi))}{Z} &= 0 \qmbox{with} \\
    \Phi V(\Phi) &= 0 \qmbox{at} Z = 0 \qmbox{and} Z = 1 \\
    \Phi(Z,T=0) &= \Phi_0 \equiv \frac{\phi_0}{\phi_m} \leq 1\,.
\end{align}
\end{subequations}
The sedimentation flux  $J (\Phi) \equiv \Phi V(\Phi)$ is zero both when $\Phi = 0$  and $\Phi = 1$, and the wave speed $U= \dd{J(\Phi)}{\Phi} = (1 -(1+k)\Phi)(1-\Phi)^{k-1}$ can be either positive or negative. We note that it is appropriate to state two boundary conditions in $Z$ despite (\ref{SedGE}a) being first order in $Z$ since the problem is hyperbolic and information may propagate into the domain at either boundary depending on the sign of $U$ \citep{whitha74_book}. In Fig. \ref{FigFlux}(b) we plot $U(\Phi)$, $J(\Phi)$ and $V(\Phi)$ according to the sedimentation model \eqref{Vnondim}, with $k = 3$. The side channel, which is located at the dimensionless distance $D = d/h$ from the base of the reservoir, continuously samples the fluid at $Z = 1 -D$, and thus may encounter a range of suspension properties over an interval of time. It is useful to define the time-averaged hematocrit $\overline{\Phi}$ at a location $Z$ during a time interval $T$ by
\begin{align}
\overline{\Phi}(Z,T) \equiv \frac{1}{T} \int_0^{T} \Phi(Z,T') \rmd T',
\end{align}
which is a useful measure of the average properties encountered at the entrance of the side channel, with $Z = 1-D$. We note that while $v_0$ is the natural velocity scale, it is an extrapolated value for the $\phi \rightarrow 0$ limit and is thus an order of magnitude greater than the sedimentation rate of typical blood $(\phi \approx 0.45)$; see Fig. \ref{FigFlux}(a).

\subsection{Numerical solutions} 
\begin{figure}[t!]
    \centering
    \includegraphics[scale=1]{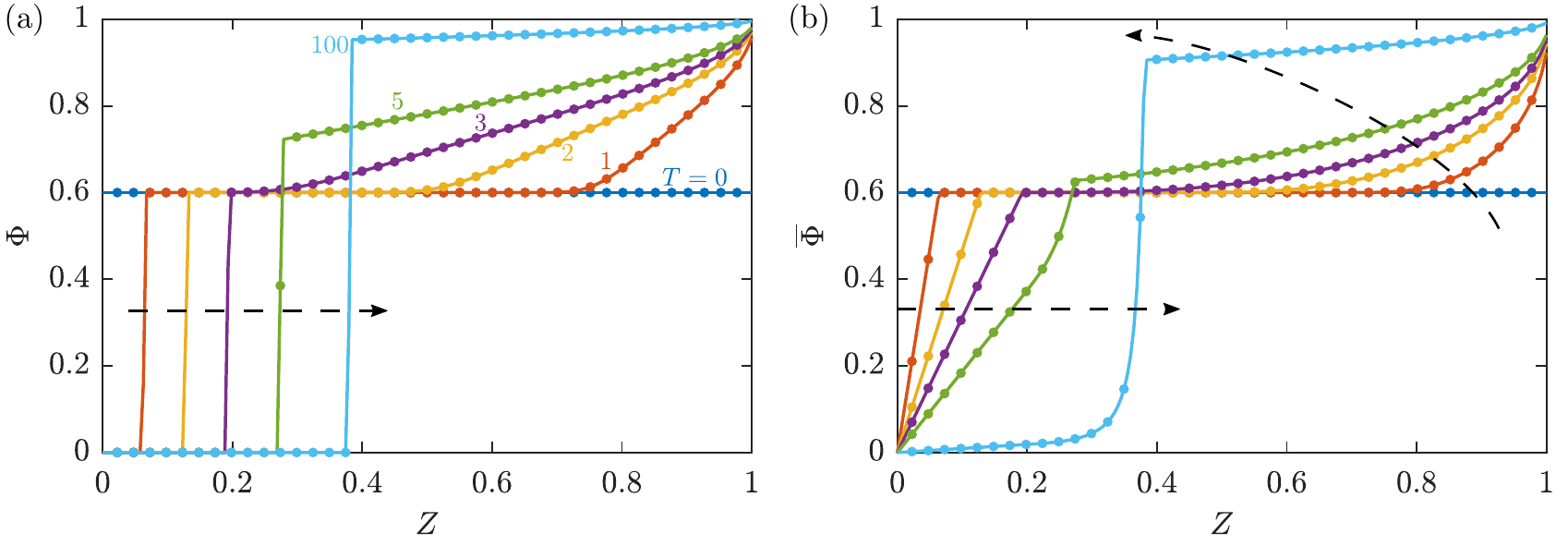}
\caption{Results of a typical simulation with initial condition $\Phi(Z,T=0) = \Phi_0 = 0.6$ ($\phi_0 = 0.45$, $\phi_m = 0.75$, $k = 3$; typical for blood) showing profiles of (a) the instantaneous hematocrit $\Phi(Z,T)$ and (b) the mean hematocrit $\overline{\Phi}(Z,T)$ as functions of the spatial coordinate $Z$, at times $T = \{0, 1, 2, 3, 5, 100\}$. Dashed arrows indicate profiles at increasing $T$. A sedimentation front propagates towards $Z = 1$ (the base), while a compression wave propagates towards $Z = 0$, until the two meet. }
\label{FigProfiles}
\end{figure}

 We solve the hyperbolic problem defined by equations (\ref{SedGE}a--c) using a first-order conservative Godunov-type scheme with upwind differencing in the direction of the local wavespeed $U$ \citep{har83_upstream}. Here and below, we use $k = 3$, consistent with experimental data (see Fig\ref{FigFlux}, \citep{rou30_sediment}). Instantaneous and time-averaged concentration profiles obtained from a typical simulation with an initially uniform concentration $\Phi_0$ are plotted in Figs. \ref{FigProfiles}(a,b), respectively. As is typical of sedimenting suspensions, the system evolves by the formation of a front with position $Z = Z_f(T)$ that propagates towards the bottom of the container (the positive $Z$ direction). Upstream of the front, the suspension contains only fluid $(\Phi = 0)$ while downstream of the front the concentration of the suspension is either greater than or equal to its initial value $\Phi_0$. Additionally, a compression wave propagates rearward (towards $Z= 0$) from the no-flux boundary at $Z = 1$ (the base of the reservoir), as shown in Fig. \ref{FigProfiles}. The concentration increases behind this compression wave (as measured in its direction of propagation), corresponding to cells settling at the bottom of the container.  Between the forward traveling sedimentation front and the rearward propagating compression wave is a ``quiet'' region where $\Phi = \Phi_0$. This region shrinks and eventually vanishes as the two waves meet (around $T = 3$ in Fig. \ref{FigProfiles}). This process determines the largest time for which the subsample is representative of the original sample, which we discuss in detail in Sec. \ref{SecSubsampling}.

The speed of the sedimentation front is given by the Rankine--Hugoniot condition
 \begin{equation}
     \dd{Z_f}{T} = \frac{\left[J\right]_{Z_f}}{\left[\Phi\right]_{Z_f}},
 \end{equation}
where $[f]_Z = f|_{Z^+} - f_{Z^-}$ is the jump of a quantity $f$ around the point $Z$. We note that in a semi-infinite system, $\Phi(Z>Z_f) = \Phi_0$ and $\Phi(Z<Z_f) = 0$, so that $\dd{Z_f}{T} = V(\Phi_0)$.  Under confinement, the system eventually reaches a steady state which, due to mass conservation, is given by $\Phi(Z,T\rightarrow\infty) = \Theta[Z-(1-\Phi_0)]$, where $\Theta$ is the Heaviside step function, with the front approaching a fixed position $Z_f \rightarrow 1-\Phi_0$ (Fig. \ref{FigProfiles}). 

\section{Subsampling the sedimenting suspension} \label{SecSubsampling}
Solutions of equation \eqref{SedGE} may be analyzed further using the method of characteristics though this is not the primary focus of the present work. Instead we are interested in sampling the sedimenting suspension -- whose properties evolve in both space and time -- in a way that is predictable and therefore of potential value to diagnostics. For applications it is often necessary that the side channel subsamples the sedimentation suspension for a prescribed sampling time $t_s$, which defines a sampling window $T \in [0, T_s \equiv \frac{t_s v_0}{h}]$. Typically, $t_s$ is set by the application and $v_0$ is a property of the suspension, with $h$ being a design parameter. Thus, we have three dimensionless parameters: the hematocrit of the sample $\Phi_0$, the sampling time $T_s$ (related to the fill height), and the location of the subsampling port $D = d/h$; $\Phi_0$ is an input  while $T_s$ and $D$ are design parameters. 

We are interested in finding the region of the design space ($T_s$, $D$) for which the side channel samples a target hematocrit to within a specified tolerance. We introduce two variants of this problem: 
\begin{enumerate}[(i)]
     \item the side channel instantaneously samples only the original sample hematocrit throughout the sampling time window, i.e. $\Phi(1-D,0<T<T_s) = \Phi_0$, and
     \item the side channel samples a target mean hematocrit $\Phi^*$ (generally $\neq \Phi_0$) over the sampling window, i.e. $\overline{\Phi}(1-D,T_s) = \Phi^*$. 
 \end{enumerate}  
 In both cases we specify a tolerance $\delta \Phi$. Note that a solution to (i) automatically solves (ii) if $\Phi^* = \Phi_0$. Due to the finite ``quiet'' region downstream of the sedimentation front as discussed above (cf. Fig. \ref{FigProfiles}), solutions to problem (i) generally result in finite regions in the ($T_s$, $D$) design space even for arbitrary small tolerance. By contrast a solution of (ii) seeking a target mean concentration $\overline{\Phi}(1-D,T_s) = \Phi^* \neq \Phi_0$ will result in solution regions whose ``size'' scales with the tolerance $\delta \Phi$.

\subsection{Representative subsampling} \label{SecRepresentative}
\begin{figure}[t!]
    \includegraphics[scale=1]{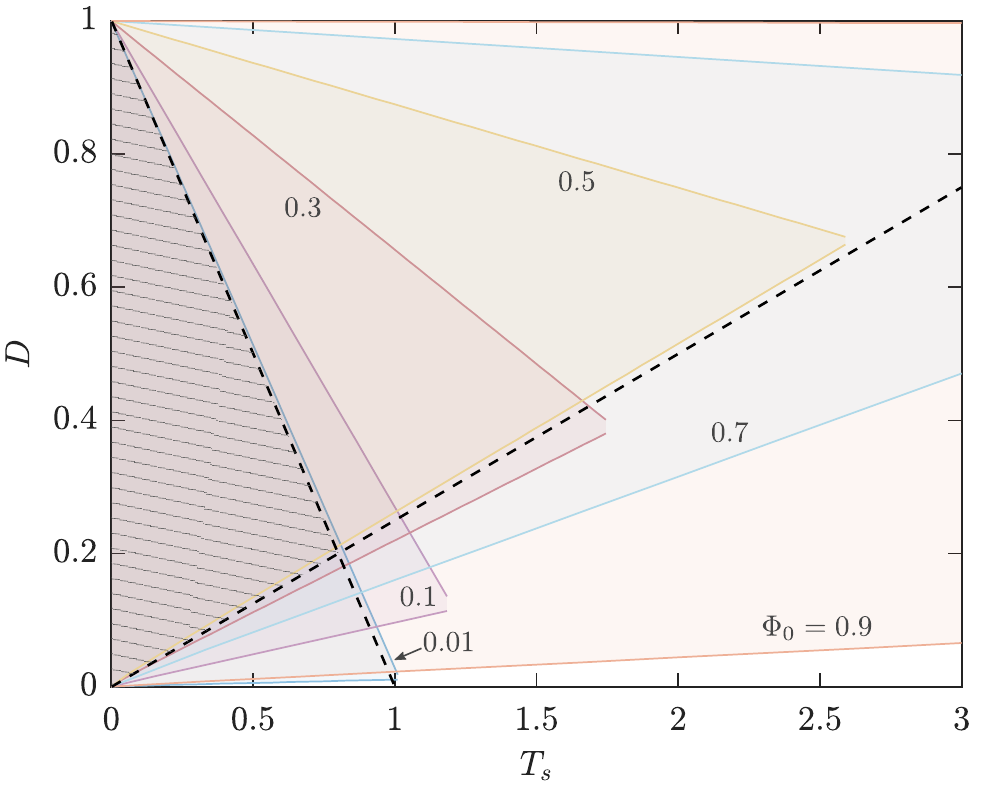}
\caption{Combinations of the dimensionless side-channel height $D$ and the dimensionless sampling time $T_s$, using $k = 3$ in \eqref{Vnondim}, for which the subsample is representative of the original sample, indicated as shaded regions for different values of $\Phi_0$, with a tolerance $\delta \Phi = 0.001$. The subsample is not representative of the original sample outside the shaded region for a particular $\Phi_0$. The regions are roughly triangular in shape. Dashed lines indicate the fastest forward and rearward propagating waves, which intersect at the point $\left(T_s = \frac{4}{5}, D = \frac{1}{5} \right)$. Along with the $T_s = 0$ axis, this point defines the triangular region of the design space (hatched) that ensures a representative subsampling for any hematocrit $\Phi_0$ [see equation \eqref{MinTriangle}].}
\label{FigSol1}
\end{figure}
We first consider the case where we demand that the subsample collected in the side channel is representative of the original sample hematocrit $\Phi_0$ throughout the sampling interval $0 < T < T_s$, corresponding to problem (i) introduced above. Thus, for a specified $\Phi_0$, we solve the system \eqref{SedGE} to obtain $\Phi(Z,T)$ and then identify combinations of sampling times $T = T_s$ and side-channel locations $Z = 1-D$ for which 
\begin{align} \label{SolCriterion1}
\Phi_0 - \delta \Phi \leq \Phi(1-D,T_s) \leq  \Phi_0 + \delta \Phi.
\end{align}
As noted above we may in principle obtain solution regions with finite size in phase space for $\delta \Phi = 0$, although doing so numerically is nontrivial. For the results below we use a small tolerance of $\delta \Phi = 0.001$. 

Admissible solutions ($T_s,\,D$) of \eqref{SedGE} and \eqref{SolCriterion1} are indicated as shaded regions in Fig. \ref{FigSol1} for different values of $\Phi_0$. These solution regions in the phase plane are of roughly triangular shape with details that depend on $\Phi_0$. In the limit of arbitrarily short sampling times, the concentration at any location $D$ is identical to the initial concentration $\Phi_0$, thus identifying one of the solution boundaries with the $T_s = 0$ axis. Furthermore, we observe that $\Phi = \Phi_0$ just downstream of the sedimentation front for short times (Fig. \ref{FigProfiles}) identifying another boundary of the triangular solution region with the trajectory of the front, $Z = V(\Phi_0) T$ or $D = 1 - V(\Phi_0) T$. The third side is set by the speed of the backward propagating compression wave [cf. Fig. \ref{FigFlux}(b)] that starts at $Z = 1$ ($D = 0$), although this speed of propagation is not trivial to obtain as a function of $\Phi_0$.

As Fig. \ref{FigSol1} shows, solution regions ($T_s,\,D$) depend strongly on the initial hematocrit $\Phi_0$. However, it may be important in an application to obtain a representative subsample of the original suspension over a range of initial concentrations $\Phi_0$. Designing such an application that is viable for any value $\Phi_0$ therefore requires us to identify the region of intersection of solutions over the entire range of $\Phi_0 \in (0,1)$. We estimate the size of this intersection region by considering the fastest sedimenting front and the fastest (rearward moving) compression wave. The fastest sedimentation wave moves with a velocity $V_{\rm max} = \mathrm{max}_{\Phi_0} \left\{V(\Phi_0)\right\} = 1$ corresponding to a trajectory $Z = T$ (or $D = 1 - T_s$; Fig. \ref{FigSol1}). The fastest rearward propagating waves are determined by the fastest negative wavespeeds in the system [cf. Fig. \ref{FigFlux}(b)]. The maximum negative wavespeed is $U_{\rm max}^{-} = U(\Phi_{\rm max}^{-}) < 0$ where $\Phi_{\rm max}^{-}$ is identified by $\dd{U(\Phi)}{\Phi} = 0$; for example $\Phi_{\rm max}^{-} = \frac{1}{2}$ and $U_{\rm max}^{-} = -\frac{1}{4}$ for $k = 3$. Thus, backward propagating waves are bounded by the trajectory $Z = 1 + U_{\rm max}^{-} T$ (or $D = -U_{\rm max}^{-} T_s$; Fig. \ref{FigSol1}). These limiting forward and backward waves intersect at $T_s = T_s^c = \left(1 - U_{\rm max}^{-}\right)^{-1}$ and $D = D^c = 1 - T_s^c$.  Thus, representative subsampling is guaranteed for any $\Phi_0$ within the triangular design-space region defined by the points 
\begin{align} \label{MinTriangle}
(T_s, D) = \left\{(0,0);\quad  (0,1) ; \quad \left(T_s^c = \frac{1}{1 - U_{\rm max}^{-}},\, D^c = \frac{-U_{\rm max}^{-}}{1 - U_{\rm max}^{-}} \right)\right\},
\end{align}
where we recall that $U_{\rm max}^{-} \leq 0$ and therefore both $D^c$ and $T_s^c$ are smaller than unity. This region is hatched in Fig. \ref{FigSol1}, showing good agreement with numerical results. Thus, placing the side channel at height $D = D^c$ allows the largest sampling time $T_s = T_s^c$ for any $\Phi_0$; for $k = 3$, $T_s = T_s^c = \frac{4}{5}$ and $D = D^c = \frac{1}{5}$. 

\begin{figure}[t!]
    \includegraphics[scale=1]{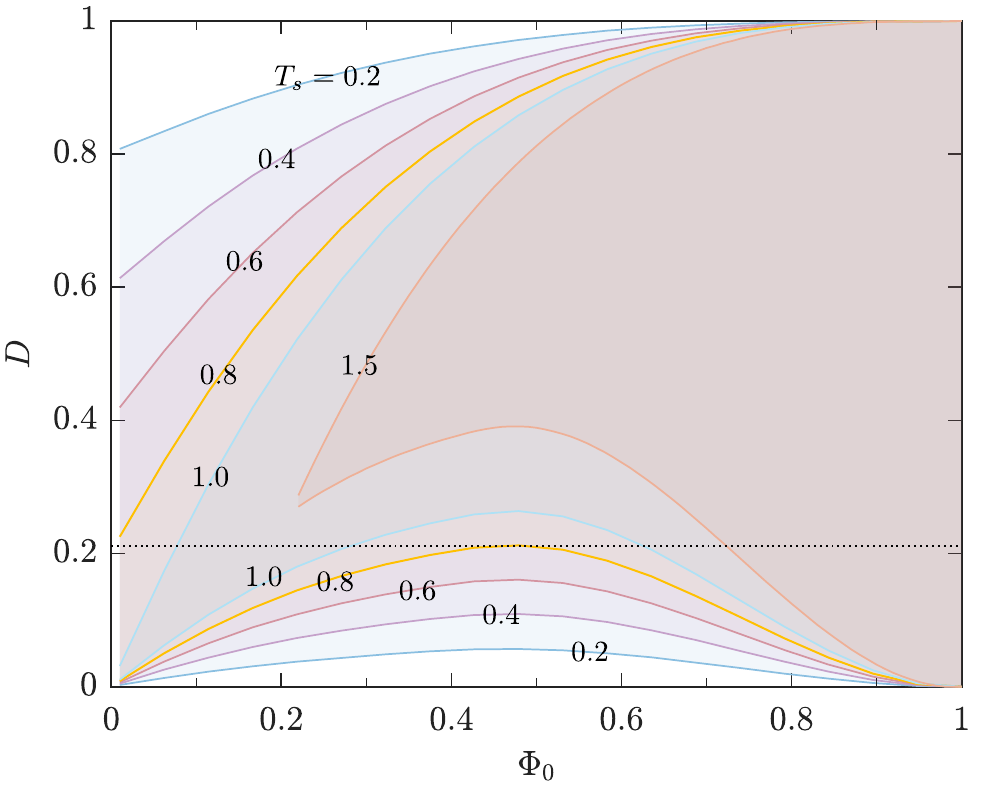}
\caption{Map of the permissible range of $D$ and sample hematocrit $\Phi_0$ for representative sampling (tolerance $\delta \Phi = 0.001$), indicated as shaded regions for different sampling times $T_s$, using $k = 3$ in \eqref{Vnondim}. The subsample is not representative of the original sample outside the shaded regions for a prescribed $T_s$. The data here are equivalent to those of Fig. \ref{FigSol1}. The dotted horizontal line represents the limiting value $D \approx 0.2$ that is marginally within the solution space for all $\Phi_0$ and occurs for $T_s \approx 0.8$, in agreement with the prediction of equation \eqref{MinTriangle}.
}
\label{FigSol2}
\end{figure}
We note that in practice the physical sampling time $t_s$ and the range of hematocrit is often prescribed so it may be necessary to choose $h$, or equivalently the dimensionless quantity $T_s = t_s v_0/h$, as a design parameter. A slightly different but equivalent visualization of Fig. \ref{FigSol1} is shown in Fig. \ref{FigSol2}, where the permissible range of $D$ is plotted against $\Phi_0$ as shaded regions for different values of $T_s$. For small $T_s$, the range of allowable $D$ for representative sampling is finite across the entire range of $\Phi_0$ values (e.g. $T_s \leq 0.6$ in Fig. \ref{FigSol2}). As $T_s$ increases up to a critical value, the range of allowable $D$ decreases to a point: at $T_s = 0.8$, only a single choice of $D \approx 0.21$ (dotted line) can guarantee a representative subsampling over all $\Phi_0$ values, consistent with the theoretical predictions $D= D^c$, $T_s=T_s^c$ in \eqref{MinTriangle}. For larger sampling times $T_s^c < T_s \leq 1$, no single value of $D$ is suitable for all $\Phi_0$ although individual solutions $D(\Phi_0)$ exist for any choice of $\Phi_0$. For even larger sampling times $T_s > 1$, representative subsampling is only possible for sufficiently large hematocrit (sufficiently slow sedimentation), as indicated by the solution for $T_s = 1.5$ in Fig. \ref{FigSol2}: no representative subsampling solutions $D(\Phi_0; T_s = 1.5)$ exist for $\Phi_0 \lesssim 0.2$.

\subsection{Oversampling and undersampling}
As we have shown above it is possible to obtain a representative subsample of the original sample, but only within a finite region of the dimensionless design space $(T_s, D)$. Outside of this solution region the collected subsample will on average differ in hematocrit from that of the original sample $\Phi_0$. This feature presents the possibility of developing a design that is capable of controlled undersampling ($\overline{\Phi}(Z,T) < \Phi_0$) or oversampling ($\overline{\Phi}(Z,T) > \Phi_0$) the suspension on average in a systematic way. Such a design may be useful to extract a more dilute or concentrated version of the original sample without the need for an extra separation or mixing step.

We denote the target mean hematocrit of the subsample by $\Phi^*$, so that we seek to design a subsampling protocol that achieves
\begin{align} \label{SolCriterion2}
\Phi^* - \delta \Phi \leq \overline{\Phi}(1-D,T_s) \leq  \Phi^* + \delta \Phi.
\end{align}
We are particularly interested in cases where $\Phi^* \neq \Phi_0$ since for $\Phi^* = \Phi_0$ (within the tolerance $\delta \Phi$) the solution to this problem is the same as that of section \ref{SecRepresentative}. 

\begin{figure}
    \includegraphics[scale=1]{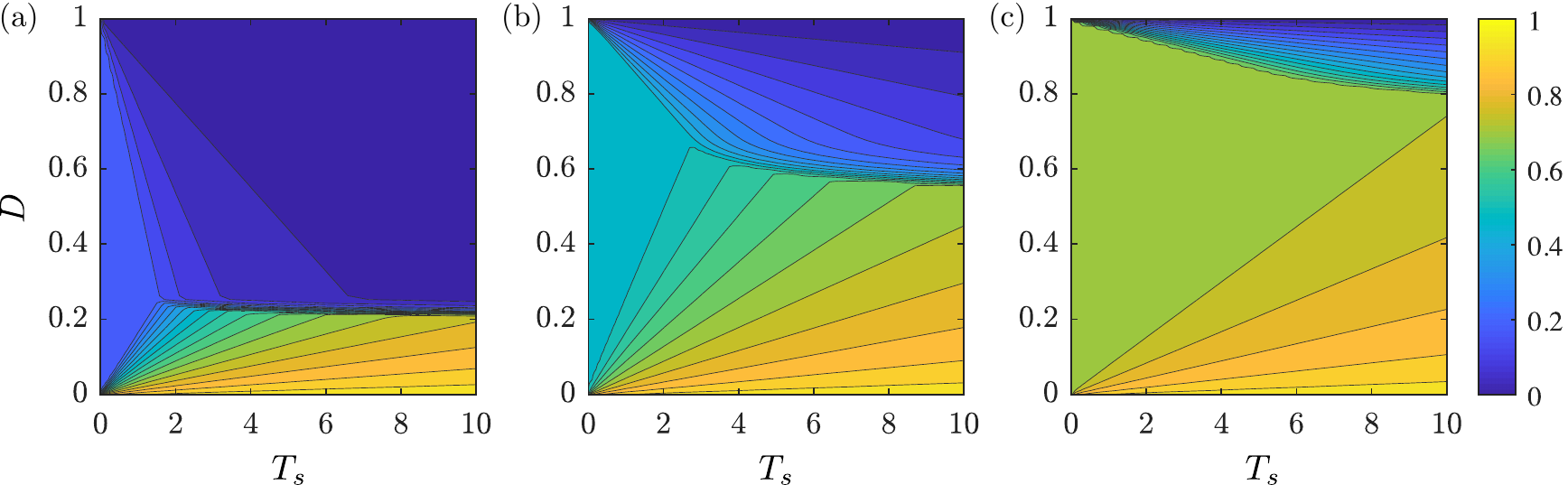}
\caption{Map of the mean hematocrit $\overline{\Phi}(1-D,T_s)$ obtained for a given side channel height $D$ and sampling time $T_s$ for different initial hematocrit values $\Phi_0$: (a) 0.2, (b) 0.5 and (c) 0.7. Contours are $\delta \Phi = 0.05$ apart. Colors correspond to $\overline{\Phi}(1-D,T_s)$, with the scale indicated by the color bar on the right. The triangular regions attached to the $T_s = 0$ axis are comparable with the shaded regions of Fig. \ref{FigSol1} and correspond to representative subsampling. The remainder of the ($D$, $T_s$) space represents either undersampling $(\overline{\Phi} < \Phi_0)$ or oversampling $(\overline{\Phi} > \Phi_0)$, corresponding to contours terminating at $D = 1$ or $D = 0$, respectively. The sample can be diluted for a larger region of the design space for small $\Phi_0$ (e.g. panel (a)) whereas the opposite is true for a large $\Phi_0$ (e.g. panel (c)).}
\label{FigSolMean}
\end{figure}

As discussed earlier in Sec. \ref{SecSubsampling}, the size of the solution region for $\Phi^* \neq \Phi_0$ will strongly depend on both $\Phi^*$ and the tolerance $\delta \Phi$. Thus, we have a 5-parameter space in general ($D, T_s, \Phi_0, \Phi^*, \delta \Phi$), which is difficult to map out in its entirely. Some insight can be obtained from the resulting map of $\overline{\Phi}(1-D, T_s)$ over the phase space $(D, T_s)$ for different initial hematocrit $\Phi_0$. Figure \ref{FigSolMean} shows contours of $\overline{\Phi}(1-D, T_s)$ for three values of the initial hematocrit $\Phi_0$ (0.2, 0.5 and 0.7). Contours are spaced $\delta \Phi = 0.05$ apart. For each panel of Fig. \ref{FigSolMean}, we note the presence of a roughly triangular region with one side along the $T_s = 0$ axis. Within the tolerance, this region corresponds with the shaded triangular regions in Fig. \ref{FigSol1}, where $\Phi(1-D,T_s)$, and therefore $\overline{\Phi}(1-D, T_s)$, is identical to the initial value $\Phi_0$. Outside of this region a range of $\overline{\Phi}$ is obtained for different combinations of $D$ and $T_s$, corresponding to the possibility of achieving different target $\Phi^*$. 

As expected, subsampling close to the top of the container $(D \approx 1)$ results in a more dilute suspension ($\overline{\Phi} < \Phi_0$), whereas subsampling near the bottom of the container $(D \approx 0)$ leads to a more concentrated suspension ($\overline{\Phi} > \Phi_0$). As $T \rightarrow \infty$ the system approaches the static solution $\Phi(Z,T) = \Theta[Z-(1-\Phi_0)]$ as discussed in Sec. \ref{SecModel}, so that the side channel samples $\overline{\Phi} \sim 0$ if $D > \Phi_0$, and $\overline{\Phi} \sim 1$ if $D < \Phi_0$. For relatively small $\Phi_0 \lesssim 0.5$ (e.g. Fig. \ref{FigSolMean}a), dilution of the subsample to a prescribed $\Phi^* < \Phi_0$ can be achieved for a larger regions of the $(T_s, D)$ design space, while precise concentration to $\Phi^* > \Phi_0$ restricts solutions to a narrower range of the design space. This behavior is reversed for a more concentrated initial suspension -- it is easier to further concentrate it (oversample) to a precise value than to dilute it (undersample) with similar precision (e.g. Fig. \ref{FigSolMean} c). For intermediate values $\Phi_0 \approx 0.5$, both dilution and concentration to a prescribed precision allow similarly sized regions of the design space (e.g. Fig. \ref{FigSolMean} b). We recall that typically $\Phi_0 \approx 0.6$ for blood, suggesting that using sedimentation it is somewhat easier to concentrate than to dilute a sample with precision. 

\section{Filling the side channel} \label{SecSideChannel}
The implicit assumption so far is that the suspension drawn into the side channel is well represented by the properties of the suspension in the reservoir at the vertical location $Z = 1-D$. However, the uptake of fluid into the side channel is itself a dynamic process that depends on several factors including the time-dependent viscosity $\eta$ of the suspension at the mouth of the side channel [a result of the time dependent hematocrit $\Phi(Z,T)$], the elapsed time and the mechanism by which fluid is pumped into the side channel. 

To illustrate this point, we assume that the fluid is driven into the side channel due to an imposed pressure drop $\Delta p$ that may be time dependent. For clarity, we denote the hematocrit in the side channel by $\psi(x,t)$, where ($x=0$, $z = h-d$) represents the mouth of the side channel where it meets the reservoir (see Fig. \ref{FigSketch}). The side channel is initially devoid of any fluid. Over time, the sample is drawn into the side channel, occupying a length $\ell(t)$ as shown in Fig. \ref{FigSketch}. For an incompressible flow, and a side channel with uniform cross-sectional area $A_s$, $\dd{\ell}{t}$ is equal to the mean fluid velocity $\overline{u}$ at any position $x \leq \ell(t)$ in the side channel.

We neglect the motion of cells relative to that of the fluid, which may be important in flows of suspensions in narrow channels and may cause particle accumulation \citep{hol11_particle_imbibition}. Instead, we assume that cells are advected passively with the flow, and that the local viscosity of the suspension depends only on the local volume fraction $\psi(x,t)$. Then,  Darcy's law for flow in the channel gives
\begin{align} \label{Darcy}
\overline{u}(t) = \dd{\ell}{t} = - \frac{\kappa}{\eta(\psi(x,t))} \pp{p}{x},
\end{align}
where $\kappa$ is the permeability of the channel and depends on its cross-sectional shape and $p(x,t)$ is the pressure. The volumetric flux through the side channel is $q = \overline{u} A_s$. 
Integrating \eqref{Darcy} yields
\begin{align} \label{DelP}
\Delta p = \frac{1}{\kappa} \dd{\ell}{t} \int_0^{\ell} \eta(\psi(x,t)) \rmd x.
\end{align}
Assuming that cells are passively advected by the mean fluid flow, the transport equation for $\psi$ in the side channel is  
\begin{align}\label{SideChannelTransport}
\pp{\psi}{t} + \dd{\ell}{t}\pp{\psi}{x} = 0,
\end{align}
with the initial condition $\psi(x = 0,t) = \phi(z = h-d,t) \equiv \psi_0(t)$. The governing equation is satisfied by a general solution of the form $\psi = f(\ell(t) - x)$, where $f$ is an arbitrary function. Using the initial condition determines $f$, so that  
\begin{align} \label{PhiSideSol}
\psi(x,t) = \psi_0\left(\ell^{-1}(\ell(t) - x)\right)
\end{align}
where $\ell^{-1}$ is the inverse function defined such that $\ell(t) = x \iff \ell^{-1}(x) = \ell^{-1}(\ell(t)) = t$. Physically, $\ell^{-1}(\ell(t) - x)$ is the time $t^*$ required to fill the side channel up to a length $\ell^* = \ell(t) - x$, i.e. $\ell(t^*) = \ell^*$. Combining \eqref{PhiSideSol} with \eqref{DelP} yields an integro-differential equation for $\ell(t)$,
\begin{align} \label{DlDtIntegroDiff}
\dd{\ell}{t} \int_0^{\ell} \eta \left[\psi_0\left(\ell^{-1}(\ell(t) - x)\right)\right)]\, \rmd x = \kappa \Delta p,
\end{align}
where we note that $\Delta p$ may be time-dependent.

\begin{figure}[t!]
    \includegraphics[scale=1]{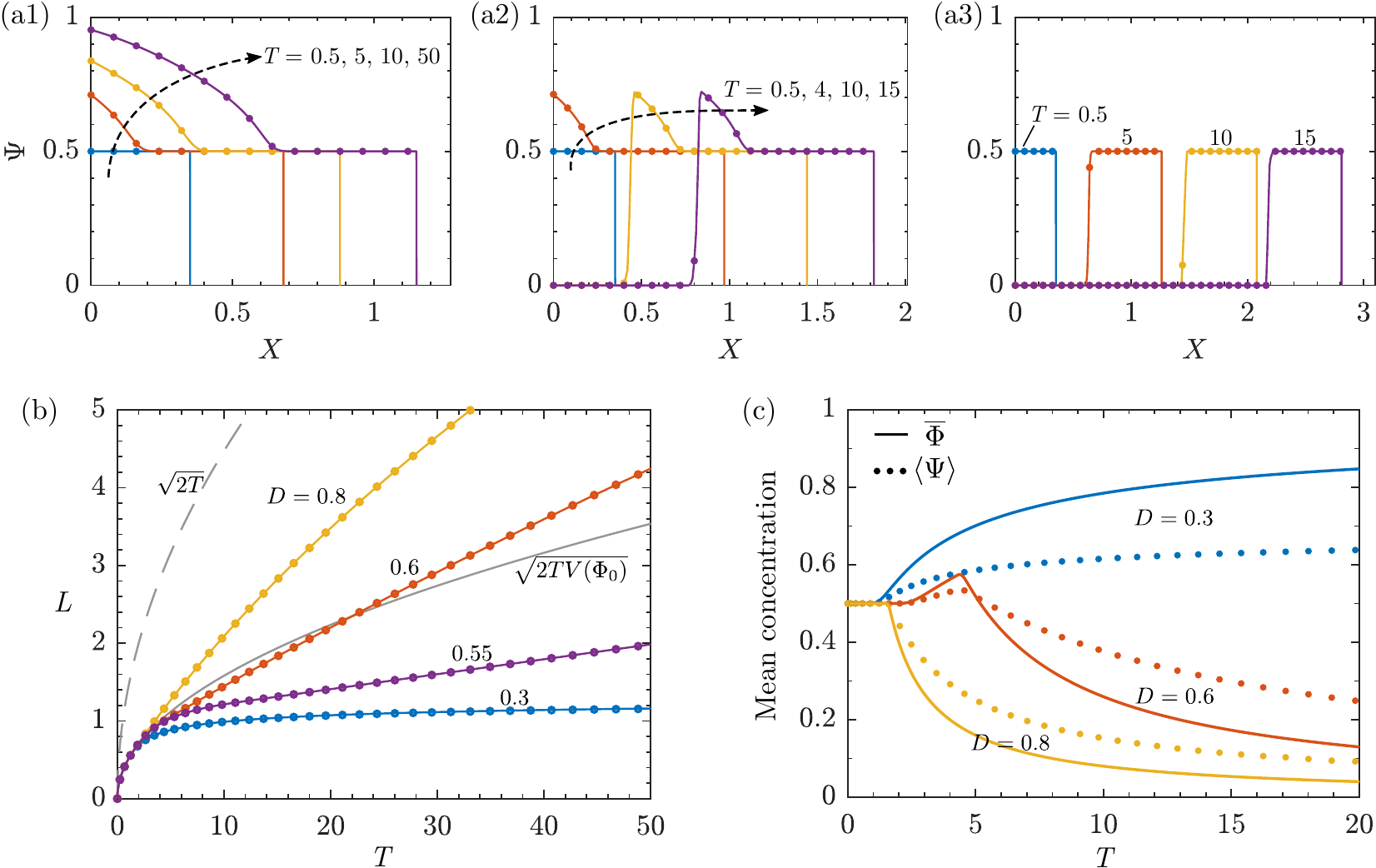}
\caption{Transport in a side channel at a constant pressure difference across the channel for $\Phi_0 = 0.5$. (a1)--(a3) Profiles of rescaled hematocrit $\Psi(X,T)$ versus $X$ for different times (indicated by dashed arrows or labels) with (a1) $D=0.3$, (a2) $D = 0.6$ and (a3) $D = 0.8$. (b) Filling length in the side channel $L(T)$ for different $D$, showing that $L(T) \sim \sqrt{2 T V(\Phi_0)}$ for small times. (c) Mean concentrations $\overline{\Phi}(1-D,T)$ (symbols) and $\left<\Psi\right> (T)$ for a constant pressure drop across the side channel (curves), plotted versus time $T$, for different channel heights $D$. Both averages are equal to each other and to $\Phi_0$ for a range of times corresponding to shaded regions in Figs. \ref{FigSol1} and \ref{FigSol2} (representative subsampling). Beyond this time $\overline{\Phi} \neq \langle \Psi\rangle \neq \Phi_0$, with $\langle \Psi\rangle$ deviating comparatively less from $\Phi_0$ than $\overline{\Phi}$. }
\label{FigSideChannel}
\end{figure}

Below, we consider the case of a constant applied $\Delta p$. We rescale time as before ($T = t v_0/h$) since the initial condition $\psi_0(t)$ has features set by sedimentation in the reservoir. We use a Krieger--Dougherty viscosity law consistent with \eqref{SedimentationLaw}, to write $\eta(\psi) = \eta_0(1 - \psi/\phi_m)^{-k}$, as discussed in Sec. \ref{SecModel}. Next, we use \eqref{DlDtIntegroDiff} to scale horizontal distances by $\ell_{\Delta p} = \left\{\kappa h \Delta p /(\eta_0 v_0) \right\}^{1/2}$ and define $X = x/\ell_{\Delta p}$ and $L(T) = \ell(t)/\ell_{\Delta p}$. Defining $\Psi(X,T) = \psi(x,t)/\phi_m$ so that $\eta(\psi) = \eta_0/V(\Psi)$ [see \eqref{Vnondim}], \eqref{DlDtIntegroDiff} rescales as 
\begin{align} \label{DlDtIntegroDiff_ND}
\dd{L}{T} \int_0^{L} \frac{1}{V \left[\Psi_0\left(L^{-1}(L(T) - X)\right)\right)]}\, \rmd X = 1,
\end{align}
where $\Psi_0(T) = \Psi(0,T) = \Phi(1-D,T)$ and $L^{-1}(L(T)) = T$. Then, the spatially averaged dimensionless hematocrit of the subsample in the side channel is 
\begin{align}
\left<\Psi\right>(T) = \frac{1}{L(T)}\int_0^{L(T)}\Psi(X,T)\,\rmd X.
\end{align}

 If $\Psi_0(T) = \Psi_0$ is constant over the time of interest, we find 
 \begin{align}
 L(T) = L_0(T) \equiv \left(2 T V(\Psi_0)\right)^{1/2}
 \end{align}
 or $\ell(t) = \left(2 t \kappa \Delta p/ \eta(\psi_0) \right)^{1/2}$, which is a result similar to capillary imbibition of wetting fluids. This result corresponds to subsampling within the shaded regions of the design space in Figs. \ref{FigSol1} and \ref{FigSol2}, for which the hematocrit at the opening of side channel ($z = h-d$, $x = 0$) is equal to that of the  original sample (representative subsampling). Consequently the subsample in the side channel is spatially homogeneous and is representative of the original sample in terms of hematocrit, i.e. $\Psi(X,T) = \left<\Psi\right>(T) = \Phi(1-D,T) = \Phi_0$.

Outside of these solution regions, we solve \eqref{DlDtIntegroDiff_ND} numerically by first-order Euler integration, using $L(T) \sim L_0(T) =\left(2 T V(\Psi_0)\right)^{1/2}$ as an asymptotic result as $T \rightarrow 0$.  In this case, the subsample is spatially inhomogeneous since the conditions presented to the inlet of the side channel change over time. Figure \ref{FigSideChannel}(a1--a3) shows profiles of $\Psi$ versus $X$ at different times and for different sampling heights. For $D< \Phi_0$ [Fig. \ref{FigSideChannel}(a1)] the concentration at the mouth of the side channel increases in time due to sedimentation in the reservoir. The opposite is true for $D> \Phi_0$ [Fig. \ref{FigSideChannel}(a2--a3)] where the side channel eventually samples $\Phi = 0$ fluid. For some values of $D \gtrsim \Phi_0$, the concentration first increases and then decreases to zero at the inlet $X =0$ [Fig. \ref{FigSideChannel}(a2)]. 

The growth of $L(T)$ is plotted in Fig. \ref{FigSideChannel}(b) for different values of $D$. In all cases $L \sim L_0(T)$ initially, with late time dynamics depending on the location $D$ of the side port. We recall the steady state sedimentation solution $\Phi(Z,T \rightarrow \infty) \rightarrow \Theta[Z - (1-\Phi_0)]$. Thus, for $D < \Phi_0$, the side channel samples increasingly more concentrated suspension, resulting in $L(T)$ approaching a constant value. For any $D > \Phi_0$, $L(T)$ grows indefinitely and may either be faster or slower than $L_0(T)$ at any finite time $T$, though it is guaranteed to be asymptotically faster than this early-time growth law as $T \rightarrow \infty$ since the side channel takes up only (low viscosity) plasma  beyond a certain time. 

It is reasonable to expect the mean hematocrit of the subsample $\left<\Psi \right>$ to be close to $\overline{\Phi}$, as discussed earlier. However, there are two sources of systematic deviation from this value. First, the flux into the side channel is greater for lower hematocrit (due to lower viscosity) resulting in a bias toward lower average cell concentrations over the entire side channel. Second, the flux is greater at earlier times when $\ell(t)$ is small (due to greater $|\pp{p}{x}| = \frac{\Delta p}{\ell(t)}$), so the mean subsample hematocrit is biased towards earlier parts of the inlet condition $\phi_0(t)$. The spatially averaged subsample hematocrit $\left<\Psi \right>$ is plotted in Fig. \ref{FigSideChannel}(c) for different $D$, showing the time average at the mouth of the side channel, $\overline{\Phi}(1-D,T)$, for comparison. As expected, the two averages are identical for sufficiently short times, for which both quantities are equal to $\Phi_0$ (representative sampling). At a critical time that depends on $D$ (see Fig. \ref{FigSol1}) both quantities deviate from $\Phi_0$ and from each other. As discussed above, we observe a tendency for $\langle \Psi\rangle$ to stay closer to the initial value $\Phi_0$ than $\overline{\Phi}$. We also note that the deviations between $\langle \Psi\rangle$ and $\overline{\Phi}$ are larger for $\overline{\Phi} > \Phi_0$, consistent with the argument that high-concentration fluid has a greater resistance to flow and therefore less of it is drawn into the channel. 

Finally, we note that in practice the pressure may be modulated so that the suspension is drawn into the side channel at constant volumetric flux (e.g. using a syringe pump). We retain rescaled variables as before, with the caveat that lengths are rescaled by a reference length $\ell_q = \frac{q h}{v_0 A_s}$ instead of $\ell_{\Delta p}$. At constant flux through a channel with constant cross-sectional area, $\dd{L}{T}$ is constant, so that \eqref{SideChannelTransport} admits the solution $\Psi(X,T) = \Psi_0\left(T\left(1-\frac{ X}{L}\right)\right)$. Consequently, the space-averaged hematocrit in the side channel is equal to the time-averaged hematocrit at its inlet, i.e. $\left<\Psi\right>(T) = \overline{\Phi}(1-D,T)$. This result for constant-flux pumping into the side channel is in contrast with the case of constant-pressure pumping, as discussed above and plotted in Fig. \ref{FigSideChannel}(c). 

Thus, the transport of the suspension within the channel adds an additional layer of complexity to sampling, over features arising from sedimentation in the reservoir, particularly when this transport is pressure-controlled. We note that the presence of viscosity gradients in the channel may produce additional flow features such as instabilities. Further, interactions of cells with channel walls or with the fluid velocity distribution may cause the cells to either lag or accumulate near the moving front in the side-channel. These effects are beyond the focus of this study but may be important in applications. 

\section{Concluding remarks} \label{SecConcl}
We have developed design considerations to subsample sedimenting suspensions, with a particular focus on blood, which may be relevant to some classes of diagnostic devices. Using a Krieger--Dougherty law along with a one-dimensional Kynch sedimention model we identify combinations of design parameters that ensure representative subsampling; these parameters relate to (i) the geometry of the container holding the original suspension, and (ii) the location of the side channel where subsampling occurs. Further, we explore the possibility of extracting a subsample that is either more or less concentrated than the original sample. We find that doing so with precision depends on several factors including the process by which fluid is pumped into the channel in which the subsample is collected. We find important distinctions between flux-controlled and pressure-controlled pumping into the side channel, particularly outside the solution region for representative sampling. 

In the present work it is implicit that the suspension may be treated as a continuum both in the reservoir and the subsampling channel. This assumption may need to be modified particularly if the side channel has transverse dimensions comparable with that of a single cell. At such scales the deformability of erythrocytes becomes an important feature. Furthermore, it has also been shown that flowing suspensions in narrow channels may result in further particle separation or instabilities, neither of which is modeled here. Nonetheless, our present work provides systematic guiding principles to design devices that extract subsamples from a larger volume of a suspension, such as blood, in ways that either exploit or compensate for sedimentation.



\begin{acknowledgments}
The authors thank Abbott Point of Care Inc. and the National Science Foundation, via grant CBET-1702693, for partial support of this work. 
\end{acknowledgments}



\end{document}